# Approximating Densities of States with Gaps

Roger Haydock

and C.M.M. Nex

Department of Physics

University of Oregon

Eugene, OR 97403-1274

USA

Reconstructing a density of states or similar distribution from moments or continued fractions is an important problem in calculating the electronic and vibrational structure of defective or non-crystalline solids. For single bands a quadratic boundary condition introduced previously [Phys. Rev. B **74**, 205121 (2006)] produces results which compare favorably with maximum entropy and even give analytic continuations of Green functions to the unphysical sheet. In this paper, the previous boundary condition is generalized to an energy-independent condition for densities with multiple bands separated by gaps. As an example it is applied to a chain of atoms with *s*, *p*, and *d* bands of different widths with different gaps between them. The results are compared with maximum entropy for different levels of approximation. Generalized hypergeometric functions associated with multiple bands satisfy the new boundary condition exactly.

1.  Boundary Conditions

The challenge in understanding extended quantum systems is that only a small part of most systems can be included in any calculation.  Connecting the properties of this small part with those of the whole depends on properties of the rest of the system.  Examples of such connections are translation symmetry in band theory, and various self-consistent or non self-consistent effective medium theories.

All these approximations can be reduced to the choice of a boundary condition applied to the states of the system at the end to the calculation.  The simplest is to fix the value of the solution to be zero on the boundary (Dirichlet), or to fix the derivative to be zero on the boundary (Neumann).  These are examples of boundary conditions which preserve time-reversal symmetry in that for real potentials, they lead to real wave functions which are singlets under time-reversal.  In physical terms they describe states which do not carry currents and are perfectly reflected at the boundary.

For crystals, the extended system can be replaced exactly by a unit cell with the periodic boundary condition that the wave functions on opposite faces of the unit cell are related by a complex phase factor.  Bloch states satisfy this boundary condition, and with the exception of singular energies, they belong to time-reversal doublets with the two states carrying currents of the same magnitude in opposite directions.

There are many extended systems such as solids with defects or disorder,

which have reduced or no periodicity, yet their states belong to bands carrying currents of various kinds, and so they belong to time-reversal doublets. Since the simple boundary conditions which fix the value of the solution or its derivative do not produce doublets, they do not reproduce bands.

In a previous paper [1], we considered several quadratic conditions designed to produce pairs of states with minimal reflections from the boundary of the system. Our approach to this problem is to construct time-reversal doublets made of states with maximally broken time-reversal symmetry (MBTS), maximal currents in opposite directions. Because of time-reversal symmetry, the Fourier components of states come in pairs, say with wave numbers $k$ and $-k$. In a time-reversal singlet, both components must occur with equal intensity and so their currents cancel. In a time-reversal doublet, the Fourier components might be separable, one to each of the states in the doublet producing the MBTS states in which magnitudes of the currents carried by each state are maximized and hence the reflections at the boundary are minimized.

Bloch states with crystal momenta $\kappa$ and $-\kappa$ have Fourier components which lie on disjoint reciprocal lattices unless $\kappa$ and $-\kappa$ are equivalent under some symmetry of the system other than time-reversal. As a result there is no cancellation and these two Bloch states form an MBTS doublet. It is not clear whether the MBTS states of non-crystalline system also have disjoint Fourier components, but it is clear that minimizing the overlap between their Fourier components maximizes the current and therefore minimizes reflections.

A powerful approach to the problem of extended systems is the method of moments which takes a variety of forms ranging from power moments, to modified moments, or, in this work, continued fraction expansions [2] which are an optimized form of modified moments. As in the discussion above, only a finite part of the infinite continued fraction can be calculated and the goal is to use this to reconstruct some distribution such as the density of states for extended systems with bands.

Many methods have been proposed for reconstructing such continuous densities from continued fractions, but each is unsatisfactory in some important respect [3]. Of the various boundary conditions studied previously, Eq. 18 of [1], stood out by producing densities of states which compare favorably with maximizing the entropy functional [4], and this boundary condition even made possible numerical continuation of the Green function to its unphysical sheet. The only problem is that this boundary condition fails for densities of states with gaps.

In this paper we generalize the above boundary condition to one which is also energy-independent and applies to densities of states with multiple bands separated by gaps. What follows is divided into five Sections, starting with the formulation of the problem as a recurrence, proceeding to the development of the generalized MBTS boundary condition, its application to a chain of atoms with three orbitals per atom, the connection of this condition to another quadratic relation, and an Appendix containing a derivation of the new boundary condition.

2. Formulation as a Recurrence

We formulate the problem of an extended quantum system as a three-term recurrence (TTR). The advantage of this approach is that any linear, Hermitian, wave-equation can be expressed in this form, not just Schrödinger's [2] and Heisenberg's [5] equations, but Maxwell's [6] equations and even Liouville's [7] equation for the classical evolution of distributions in phase space, can be transformed into a TTR. In this approach, the state of energy $z$ is $\{\psi_n(z)\}$ which satisfies the TTR,

$$\psi_{n+1}(z) = (z - a_n)\, \psi_n(z) - \beta_n\, \psi_{n-1}(z), \qquad (1)$$

where the parameters $\{a_n\}$ and $\{\beta_n\}$ are determined from the construction of a basis $\{U_0, U_1, U_2, \ldots, U_n, \ldots\}$ of states which tridiagonalize the Hamiltonian $H$ (or corresponding operator for other equations of motion) producing the related recurrence,

$$U_{n+1} = (H - a_n)\, U_n - \beta_n\, U_{n-1}. \qquad (2)$$

The above monic form of the TTR for $\{\psi_n(z)\}$ and $\{U_n\}$ is used here because its solution can be expressed in theta functions [see the Appendix]. It differs from the usual symmetric form of the recurrences only in the normalizations of the $\{U_n\}$ and

$\{\psi_n(z)\}$. The first element of the basis, $U_0$ is special in that only states contained in $U_0$ are spanned by the basis.

A finite calculation produces only a finite number of the parameters, say $a_0$, $a_1$, ..., $a_{N-1}$, and $\beta_1$, $\beta_2$, ..., $\beta_{N-1}$. In these terms the desired boundary condition is some relation between $\psi_{N-L}(z)$, $\psi_{N-L+1}(z)$, ..., $\psi_{N-1}(z)$, $\psi_N(z)$. For the single-band case, this relation[1] is the quadratic,

$$\psi_N(z)\,\psi_{N-2}(z) = \psi_{N-1}(z)^2. \tag{3}$$

For multiple bands with certain symmetries, the recurrence parameters approach a limit with period M, $a_{n+M}=a_n$ and $\beta_{n+M}=\beta_n$, then $\psi_N(z)/\psi_{N-m}(z)=\psi_{N-M}(z)/\psi_{N-M-m}(z)$ for any $0<m<M$. However, m=1 requires calculation of the fewest components of the solution, just $\psi_{N-M-1}(z)$, ..., $\psi_N(z)$, so the generalization [1] of Eq. 3 for these multiple symmetric bands is taken to be,

$$\psi_N(z)\,\psi_{N-M-1}(z) = \psi_{N-1}(z)\,\psi_{N-M}(z), \tag{4}$$

which includes the single-band case when M=1.

When there are multiple bands, the recurrence parameters do not generally approach a periodic limit. Instead the approach a limit which is almost periodic, a class of recurrences which displays a rich variety of behaviors: for example Harper's equation [8] which has $\beta_n = 1$ and $a_n = a\,cos(nx)$ and produces a devil's staircase spectrum when $x/\pi$ is irrational. The problem addressed in the next sections is the development of a boundary condition for MBTS states of these recurrences and its application to a density of states containing three bands.

3. A Generalized Quadratic Boundary Condition

The success of the quadratic boundary condition for a single band and its extension to symmetric bands raises the question of how it can be generalized to arbitrary multiple bands. Some of the possibilities include equations of higher degree such as cubic or quartic, or energy-dependence in the terms of the boundary condition. What is clear is that for more bands, more values of $\{\psi_n(z)\}$ will be needed than in the above cases.

It seems reasonable to expect a generalized boundary condition to be quadratic like the above conditions because its two solutions should be the two MBTS states of a doublet, except at singular energies where the quadratic should become degenerate. Furthermore, a generalization must include the boundary conditions for the single band and periodic recurrences as special cases. This restricts the generalization to linear combinations of terms of the form $\psi_{N-k}(z)\psi_{N-L+k}(z)$ for $0 \leq k \leq L/2$.

Surprisingly, solutions of the almost periodic TTR which is the limit of the TTR for a density with multiple bands, exactly satisfy the following quadratic relation with coefficients independent of $z$, as is shown in the Appendix. Depending on whether L is even, equal to 2K, or odd, equal to 2K+1, the boundary conditions are respectively:

$$c_{N,0}\, \psi_N(z)\, \psi_{N-2K}(z) + c_{N,1}\, \psi_{N-1}(z)\, \psi_{N-2K+1}(z) + \ldots$$

$$+ c_{N,K-1}\, \psi_{N-K+1}(z)\, \psi_{N-K-1}(z) + c_{N,K}\, \psi_{N-K}(z)^2 = 0, \qquad (5a)$$

or,

$$c_{N,0}\, \psi_N(z)\, \psi_{N-2K-1}(z) + c_{N,1}\, \psi_{N-1}(z)\, \psi_{N-2K}(z) + \ldots$$

$$+ c_{N,K}\, \psi_{N-K}(z)\, \psi_{N-K-1}(z) = 0. \qquad (5b)$$

Just as in the one band case, this relation seems also to give good MBTS solutions for TTRs which approach an almost periodic limit, as well as for those which are already almost periodic. The example in the next Sec. illustrates this useful property of the relation.

Because the coefficients $\{c_{N,k}\}$ in the above relation are independent of $z$, they apply for values of complex $z$ far from the spectrum of the recurrence. In this part of the complex energy-plane, the divergent MBTS solution dominates Eq. 1 for any initial conditions. Different choices of such $z$ provide a set of homogeneous linear equations for the coefficients $\{c_{N,k}\}$, which turn out to be real because the recurrence is Hermitian, and add up to zero because a pure exponential must satisfy the relations.

The even case, Eq. 5a, with K=1 gives the boundary condition for a single band, or TTR of period one, while the odd case, Eq. 5b, with K=1 gives the condition for a TTR of period 2, two symmetric bands. More generally, the condition for a TTR of period M=2K-1, odd, is a special case of Eq. 5a with $c_{N,0}=c_{N,1}=1$ and the rest zero, or a TTR of period M=2K, even, is a special case of Eq. 5b again with $c_{N,0}=c_{N,1}=1$ and the rest zero.

Given $a_0, a_1, \ldots, a_{N-1}$, and $\beta_1, \beta_2, \ldots, \beta_{N-1}$, it is conventional to construct polynomials $P_1(z), P_2(z), \ldots, P_N(z)$, and $Q_2(z), Q_3(z), \ldots, Q_N(z)$ which are solutions to Eq. 1 with the initial conditions $P_{-1}(z)=Q_0(z)=0$ and $P_0(z)=Q_1(z)=1$. Since the recurrence has only two linearly independent solutions for each $z$, the MBTS solution can be written as a linear combination of the polynomials,

$$\psi_n(z) = P_n(z) R(z) - Q_n(z), \tag{6}$$

where the projected density of states (PDoS) $n_0(E)$ conveniently turns out to be [2],

$$n_0(E) = |Im\{R(E)/\pi\}|, \tag{7}$$

for real energies $E$. The problem is that the PDoS depends on an MBTS solution to the recurrence which is some unknown combination of the two polynomial solutions. The importance of the boundary condition in Eq. 5 is that it determines the combination of polynomials which are MBTS which in turn determines the PDoS. However, for K asymmetric bands, the K+1 coefficients $c_{N,0}$, $c_{N,1}$, ..., $c_{N,K}$ in Eq. 5a, must first be calculated.

For $z$ far from any band, the MBTS solution $\psi_n(z)$ for the physical sheet increases exponentially with n and the one for the unphysical sheet decreases exponentially. The consequence of this is that for such $z$, $P_n(z)$ and $Q_n(z)$ both converge at least exponentially to one of the MBTS solutions. Because the $\{c_{N,k}\}$ are independent of $z$, they can be calculated far from the bands and applied near the bands and on the unphysical sheet. This is implemented numerically by constructing polynomials for values of $z$ far from any band, and since the $\{c_{N,k}\}$ are real, each complex value of $z$ gives two linear relations for the $\{c_{N,k}\}$ while each real value of $z$ gives one linear relation. K-1 such relations are needed along with the two conditions, that all the $\{c_{N,k}\}$ sum to zero and that one of them may be set to one (normalization), in order to determine the K+1 constants $\{c_{N,k}\}$.

Once the $\{c_{N,k}\}$ are known, Eq. 6 can be substituted into Eq. 5 to give a quadratic equation for $R(E)$,

$$A_N(E)\, R(E)^2 - B_N(E)\, R(E) + C_N(E) = 0, \tag{8}$$

where,

$$A_N(E) = \sum c_{N,k}\, P_{N-k}(E)\, P_{N-L+k}(E), \tag{9}$$

$$B_N(E) = \sum c_{N,k}\, [P_{N-k}(E)\, Q_{N-L+k}(E) + P_{N-L+k}(E)\, Q_{N-k}(E)], \tag{10}$$

and,

$$C_N(E) = \sum c_{N,k}\, Q_{N-k}(E)\, Q_{N-L+k}(E), \tag{11}$$

and all the sums are over k from 0 to K. Note that the sum rule for the $\{c_{n,m}\}$ causes the leading term in each of Eqs. 9-11 to cancel so that $A_N(E)$ is of degree 2N-1, $B_N(E)$ is of degree 2N-2, and $C_N(E)$ is of degree 2N-2. The solution to the quadratic in Eq. 8 has square-root branch points at the zeros of $B_N(E)^2 - 4A_N(E)C_N(E)$ which must all be real. The analytic properties of the $R(E)$ [2] require that successive pairs of these branch points be connected by cuts along the real $E$-axis. Using Eqs. 7 and 8, it is convenient to write the MBTS density of states in the form,

$$n_0(E) = \{[B_N(E)^2 - 4 A_N(E)\, C_N(E)] / A_N(E)^2\}^{\frac{1}{2}} /(2\pi), \tag{12}$$

where all the polynomials are taken inside the square-root in order minimize the effects of near cancellations, a point discussed further in Sec. 4.

4. A Chain of Atoms with Three Bands

As an example of the application of MBTS using the extended condition, we present in this Sec. approximate calculations of the PDoS for a semi-infinite chain of atoms. For a basis we take $s$, $p$, and $d$- Wannier orbitals [9] on each atom and project the density of states on the first atom of the chain with weights 1/6, 1/3, and 1/2, respectively on the three orbitals of that atom. Taking an arbitrary energy scale, the $d$-band is centered at -11/6 with a matrix-element 1/12 between $d$-orbitals on neighboring atoms; the $p$-band is centered at -5/6 with a matrix-element 1/6 between neighbor $p$-orbitals; and the $s$-band is centered at 3/2 with matrix-element 1/4. For this example, a chain of atoms has the advantage that there are no singularities in the spectrum other than at band-edges.

Tridiagonalizing the Hamiltonian for this chain of atoms does not produce an almost periodic TTR as is used in the Appendix, but instead it generates a TTR which has such an almost periodic TTR as its limit. The parameters of the TTR are plotted in Fig. 1, illustrating the complexity produced by the three bands. While this example is too complicated to see the almost periodic limit of the parameters, the effect can be seen in the examples of a single gap in Ref. 3.

The scheme of approximation, described in the previous Sec., was implemented by generating $\{P_n(z)\}$ for $z$= -10+10$i$, 10$i$, and 10+10$i$, and these were used to construct $\{c_{N,k}\}$ for N=6, 11, 14, 16, and k=0, 1, 2, 3. Note the index of the last coefficient which must be calculated for a given N is one fewer than N, so these should be compared with approximations of what is frequently called 5, 10, 13, and 15 levels.

The MBTS results are compared in Fig. 2 with the results of maximizing the entropy functional [4] constrained by the same parameters as the MBTS, and in each case plotted at intervals of 0.01 on the *E*-axis. Maximum entropy is used here because it is 'unbiased' in the sense of making no assumptions about the moments beyond those calculated. The maximum entropy density of states is simply the exponential of a real polynomial of degree equal to twice the number of recurrence parameters [10]. Fitting this exponential to the parameters is highly non-linear, so as the number of parameters increase, the fit becomes difficult to find. In this case 27 parameters was the largest problem for which the program used here [11] was able to find a fit. For recent work on this optimization, see Ref. 12.

The behavior of the MBTS approximation is very different from the maximum entropy. With increasing N, it converges rapidly at most energies with the errors concentrated at a few energies in the form of a spurious gap or glitch. This phenomenon is reminiscent of the way tridiagonalization of a matrix [13] converges individual eigenvalues, one at a time. One interpretation of this example is that the MBTS converges a band at a time starting with the narrowest and finishing with the widest. Another way to interpret the MBTS results is as a kind of self-consistent approximation because of the solution of the quadratic. Self-consistency frequently does well in large parts of a parameter space, but poorly in some small regions. Here, for small N there are spurious gaps which narrow and disappear with increasing N as is discussed in the following Sec.

The rate of convergence in the MBTS approximation is controlled by the nature of the singularities as well as the number. Square-roots or inverse square-roots converge rapidly, while discontinuities and logarithmic divergences converge slowly [1], with the rate of convergence inversely proportional to the number of each kind of singularity.

5. Another Quadratic Relation

In addition to the quadratic relation in Eq. 8, an ideal $R(E)$ satisfies another quadratic with polynomial coefficients and the purpose of this Sec. is to relate the two quadratics. Recalling that the imaginary part of $R(E)$ is proportional to the PDoS, Ref. 3 (Eqs. 4.45 and 4.46) presents expressions for the two sheets of an ideal $R(E)$ generated by the recurrence used in the Appendix. These expressions can be written as the solutions of the quadratic equation,

$$\beta F(z) R(z)^2 - W(z) R(z) + G(z) = 0, \qquad (13)$$

where $F(z)$ is a monic polynomial of degree K-1 with one zero in each gap, $W(z)$ is a monic polynomial of degree M, G(z) is another polynomial with properties similar to F(z), and $\beta$ is a positive real number. The discriminant of Eq. 13 is the monic polynomial of degree 2K,

$$X(z) = W(z)^2 - 4 \beta F(z) G(z), \qquad (14)$$

which has one zero at each of the 2K band edges.

Equation 13 is very different from Eq. 8 in that the coefficients of R(z) in Eq. 13 are of degrees K-1, K, and K-1 respectively, while those of R(E) in Eq. 8 are of higher degrees depending on N. Since both equations give the same answer, the coefficients must be proportional to one another, and this can only occur by some combination of cancellations, additive or multiplicative. Additive cancellations occur when the sums in Eqs. 9-11 produce coefficients of zero for various powers of $E$, and multiplicative cancellations occur when the coefficients defined in Eqs. 9-11 contain common factors.

The evidence from the numerical example suggests some hypotheses about these cancellations. The first is that the glitches within the bands for N small are much more likely to be due to error in multiplicative than additive cancellation because small contributions from high powers of $E$ should produce spurious zeros near infinite energy, not in the bands. Since the expression for the density of states depends on ratios of the coefficients in the quadratic, small discrepancies in the zeros of coefficients would produce just what is seen, namely narrow features bounded by a zero and a pole – the pair which should have canceled.

It is reasonable to suppose that as the number of levels in the MBTS approximation increases, the errors in the multiplicative cancellations should decrease, making the glitches in the bands narrower, but not eliminating them. In Fig. 2 these glitches seem to disappear with increasing N, but this is probably because the interval between energies plotted remains constant while the glitches get narrower. This suggests a crude way of eliminating the glitches, namely plotting on a sufficiently course grid. A better approach would be to impose cancellation between zeros which are closer than some estimate based on the number of levels.

Acknowledgments

The Authors are grateful for support from the Richmond F. Snyder Fund, and thank A. Magnus for advice and encouragement. RH acknowledges the generous hospitality of the Theory of Condensed Matter Group, Cavendish Laboratory, Cambridge, and Pembroke College, Cambridge.

Appendix – Derivation of the Quadratic Relation

Suppose the solutions $\{\psi_n(z)\}$ of a TTR satisfy the product rule, that for some N, M, and k=0, 1, …, K the products $\psi_{N-k}(z)\ \psi_{M+k}(z)$ all lie within the same K-dimensional space of functions of z, spanned by the functions $\{\varphi_1(z), \ldots \varphi_k(z), \ldots, \varphi_K(z)\}$. It follows that there are K+1 coefficients $\{C_k\}$ such that,

$$\sum C_k\ \psi_{N-k}(z)\ \psi_{M+k}(z) = 0, \qquad (A1)$$

where the sum is over k from 0 to K. It now remains to show that solutions to the TTR for multiple bands satisfy the above product rule.

Chebyshev polynomials are an example of polynomials orthogonal on a single interval. The generalization of this problem to several bands or intervals has been studied by Akhiezer [14], Magnus [15], and Chen and Lawrence [16] among others. The construction of the polynomials proceeds by introducing an ideal weight distribution (PDoS) on the intervals, derived from a special case of the Eq, 13 in which the K-1 zeros of F(z) are at the leading band edge in each gap. The MBTS solutions needed for this work are an intermediate step in the construction of the orthogonal polynomials. It is shown below that these solutions satisfy the above product rule.

The building blocks for MBTS solutions to these ideal M-band recurrences

are theta functions:

$$\theta(v;B) = \sum \exp\{i\pi \boldsymbol{\ell} \cdot B \boldsymbol{\ell} + 2\pi i \boldsymbol{\ell} \cdot v\}, \tag{A2}$$

where the sum is over an K-1-dimensional integer lattice $\boldsymbol{\ell}$, $B$ is $i$ times a positive symmetric real K-1 by K-1 matrix, and $v$ is an K-1-dimensional vector with complex components. In other words, $\theta(v;B)$ is a K-1-dimensional Bloch wave with Gaussian components. From Ref. 15 (p4661 Eq. 4.22, and p4663 Eq. 5.12), the n-dependence of the MBTS solutions to the K-band recurrence is,

$$\psi_n(z) = \omega^n \, \theta(v' + n\boldsymbol{\delta}';B) \, / \, \theta(\boldsymbol{\eta}' + n\boldsymbol{\delta}';B), \tag{A3}$$

where $v'$ and $\omega$ depend on $z$, but $B$, $\boldsymbol{\eta}'$, and $\boldsymbol{\delta}'$ do not.

Because the above expression for $\{\psi_n(z)\}$ contains a ratio of theta functions with the same matrix $B$, the arguments can be rotated in the K-1-dimensional space, and their imaginary parts rescaled to obtain,

$$\psi_n(z) = \omega^n \, \theta(v + n\boldsymbol{\delta}) \, / \, \theta(\boldsymbol{\eta} + n\boldsymbol{\delta}), \tag{A4}$$

where $\theta(v)$ is $\theta(v;iI)$ for the identity $I$, while $v$, $\boldsymbol{\eta}$ and $\boldsymbol{\delta}$ are the rotated and rescaled versions of $v'$, $\boldsymbol{\eta}'$, and $\boldsymbol{\delta}'$. The only factors in Eq, A3 which depend on $z$ are $\omega^n$ and

θ(**v** + n**δ**).

The next step is to show that the second factor θ(**v** + n**δ**) satisfies the product rule using the definition of the theta function, Eq. A2,

$$\theta(x+y)\,\theta(x-y) =$$

$$\sum\sum [\exp\{i\pi(\mathbf{k}+\boldsymbol{\ell})^2/2 + 2\pi i(\mathbf{k}+\boldsymbol{\ell})\cdot x\}\,\exp\{i\pi(\mathbf{k}-\boldsymbol{\ell})^2/2 + 2\pi i\,(\mathbf{k}-\boldsymbol{\ell})\cdot y\}], \qquad (A5)$$

where the sums are over K-1-dimensional integer lattices indexed by **k** and **ℓ**. The sums of the products can be expressed as products of sums noting that when a particular component of **k**+**ℓ** is even (odd), the corresponding element of **k**-**ℓ** is also even (odd). Because θ(**v**) has hypercubic symmetry in the K-1-dimensional space of **v**, it doesn't matter which components of the lattice sums are even, only how many, so

$$\theta(x+y)\,\theta(x-y) = \sum (M-1)!\,\varphi_k(x)\,\varphi_k(y)\,/[k!\,(M-k-1)!], \qquad (A6)$$

where,

$$\varphi_k(v) = \sum \exp\{i\pi(\boldsymbol{\ell}_k)^2/2 + 2\pi i\boldsymbol{\ell}_k\cdot v\}, \qquad (A7)$$

and the sum is over the sub-lattice of the integer lattice, with k even and K-k-1 odd components. This is the desired result showing that the theta functions satisfy the product rule. Multiplying both sides of Eq. A6 by $\omega(z)^{2n}$ in order to recover the $z$-dependence of $\psi_{n+m}(z) \psi_{n-m}(z)$ does not affect the product rule and so the solutions to the Akhiezer TTR satisfy the product rule and hence the boundary condition.

Figure Captions

Figure 1: A plot of the parameters of the TTR for the chain of atoms in the example. Full marks indicate the values of $\{a_n\}$ and refer to the axis on the left, while the outlined marks indicate values of $\{\beta_n\}$ and refer to the axis on the right.

Figure 2: A comparison of the generalized MBTS (full line) and maximum entropy [4] (dotted line) approximations for a three-band model for N=6, 11, 13, and 16 (see text) corresponding to 11, 21, 27, and 31 moments for curves (a), (b), (c), and (d) respectively. Successive curves are displaced vertically by -0.5, and curve (d) can be taken as exact.

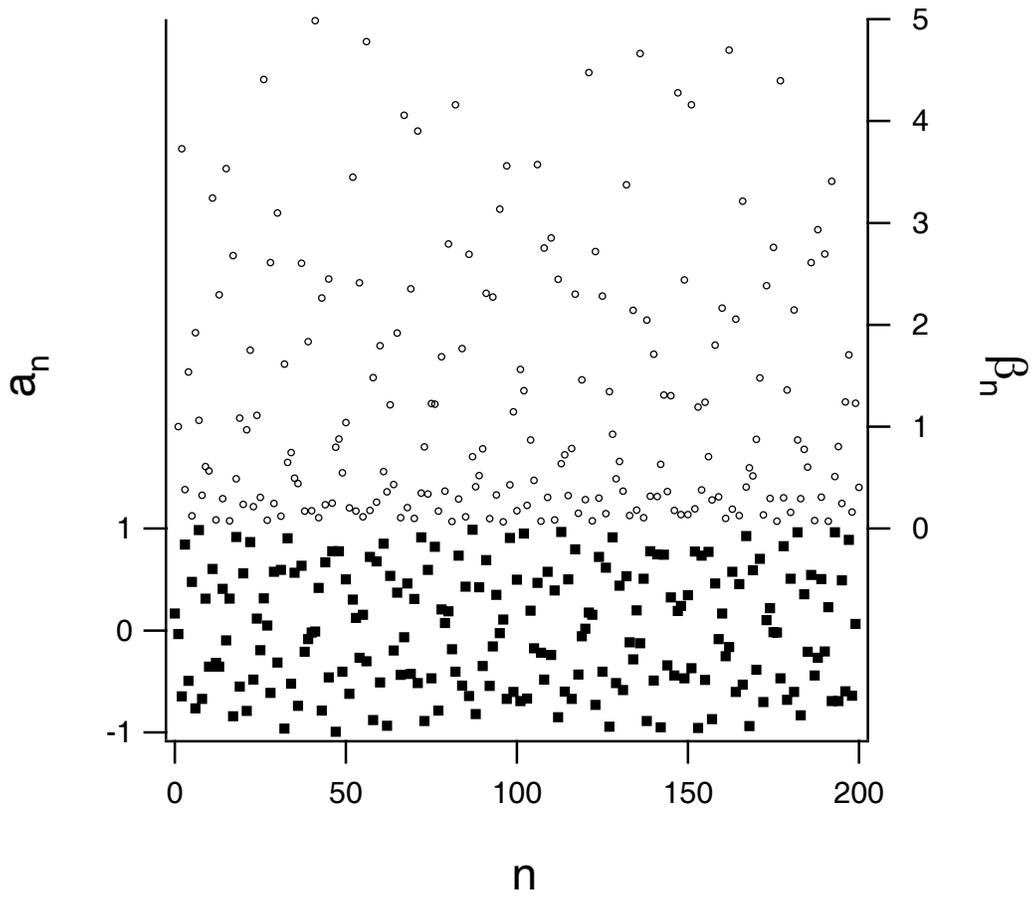

Figure 1

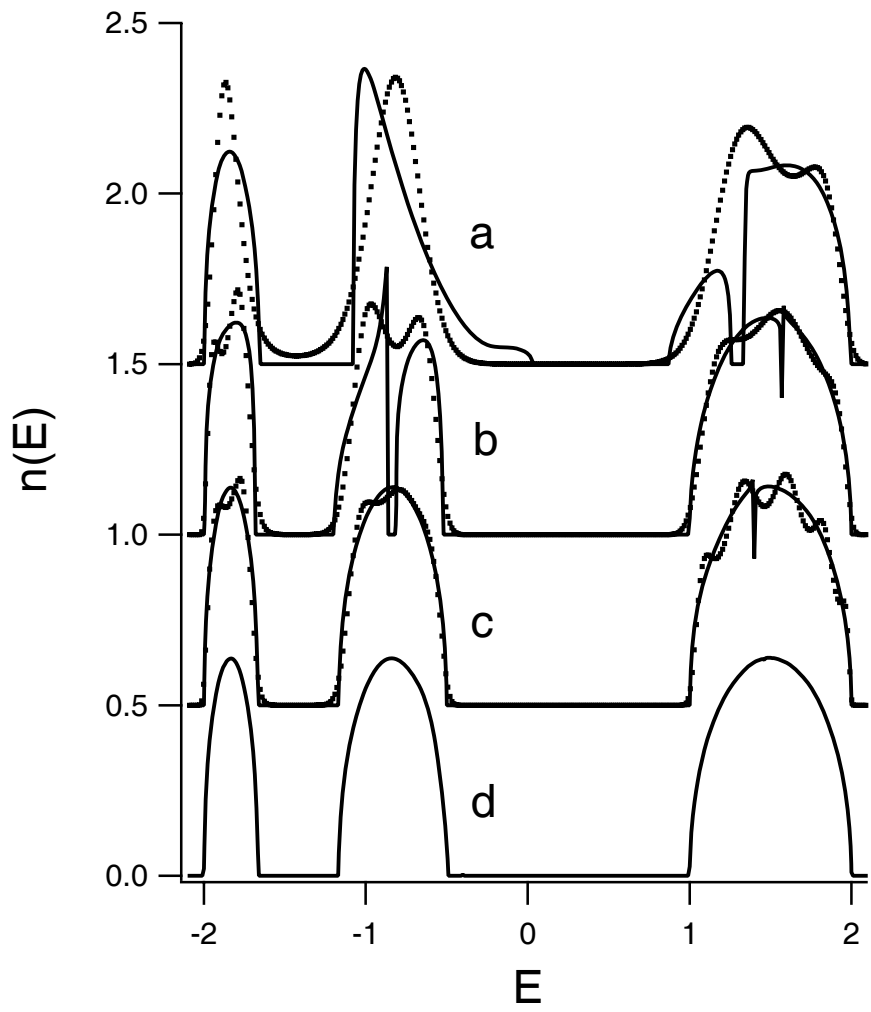

Figure 2